


\input harvmac

\def\boringfonts{y}  


\def\fonttest{y}
\ifx\boringfonts\fonttest\else

\fi

\hyphenation{anom-aly anom-alies coun-ter-term coun-ter-terms
dif-feo-mor-phism dif-fer-en-tial super-dif-fer-en-tial dif-fer-en-tials
super-dif-fer-en-tials reparam-etrize param-etrize reparam-etriza-tion}


%
%
%
\newwrite\tocfile\global\newcount\tocno\global\tocno=1
\ifx\bigans\answ \def\tocline#1{\hbox to 320pt{\hbox to 45pt{}#1}}
\else\def\tocline#1{\line{#1}}\fi
\def\toclead{\leaders\hbox to 1em{\hss.\hss}\hfill}
\def\tnewsec#1#2{\xdef #1{\the\secno}\newsec{#2}
\ifnum\tocno=1\immediate\openout\tocfile=toc.tmp\fi\global\advance\tocno
by1
{\let\the=0\edef\next{\write\tocfile{\medskip\tocline{\secsym\ #2\toclead\the
\count0}\smallskip}}\next}
}
\def\tnewsubsec#1#2{\xdef #1{\the\secno.\the\subsecno}\subsec{#2}
\ifnum\tocno=1\immediate\openout\tocfile=toc.tmp\fi\global\advance\tocno
by1
{\let\the=0\edef\next{\write\tocfile{\tocline{ \ \secsym\the\subsecno\
#2\toclead\the\count0}}}\next}
}
\def\tappendix#1#2#3{\xdef #1{#2.}\appendix{#2}{#3}
\ifnum\tocno=1\immediate\openout\tocfile=toc.tmp\fi\global\advance\tocno
by1
{\let\the=0\edef\next{\write\tocfile{\tocline{ \ #2.
#3\toclead\the\count0}}}\next}
}
%
%

%
%
%
%

\gdef\prlmode{F}
\long\def\optional#1{}
\def\cmp#1{#1}         
%
%
\let\narrowequiv=\equiv
\def\equiv{\;\narrowequiv\;}

\fontdimen16\tensy=2.7pt\fontdimen17\tensy=2.7pt 



%

%
%

%
%
%
\def\boxit#1#2{
        $$\vcenter{\vbox{\hrule\hbox{\vrule\kern3pt\vbox{\kern3pt
	\hbox to #1truein{\hsize=#1truein\vbox{#2}}\kern3pt}\kern3pt\vrule}
        \hrule}}$$
}


%




\def\splitexact#1#2{\mathrel{\mathop{\null{
\lower4pt\hbox{$\rightarrow$}\atop\raise4pt\hbox{$\leftarrow$}}}\limits
^{#1}_{#2}}}

%
%
\def\pa{\partial}

%
%
%
%

\def\dd{\mskip 1.3mu{\rm d}\mskip .7mu} 



%
%

\def\IM{isomorphism}

%
%

\ifx\boringfonts\fonttest
\font\blackboard=cmssbx10 \font\blackboards=cmssbx10 at 7pt  
\font\blackboardss=cmssbx10 at 5pt
\else
\font\blackboard=msym10 \font\blackboards=msym7   
\font\blackboardss=msym5
\fi
\newfam\black
\textfont\black=\blackboard
\scriptfont\black=\blackboards
\scriptscriptfont\black=\blackboardss


%
\ifx\boringfonts\fonttest
\font\gothic=cmssbx10 \font\gothics=cmssbx10 at 7pt  
\font\gothicss=cmssbx10 at 5pt
\else
\font\gothic=eufm10 \font\gothics=eufm7
\font\gothicss=eufm5
\fi
\newfam\gothi
\textfont\gothi=\gothic
\scriptfont\gothi=\gothics
\scriptscriptfont\gothi=\gothicss

{\catcode`\@=11\gdef\oldcal{\fam\tw@}}
\newfam\curly
\ifx\boringfonts\fonttest\else
\font\curlyfont=eusm10 \font\curlyfonts=eusm7
\font\curlyfontss=eusm5
\textfont\curly=\curlyfont
\scriptfont\curly=\curlyfonts
\scriptscriptfont\curly=\curlyfontss
\def\cal{\fam\curly\relax}
\fi
%

\ifx\boringfonts\fonttest\else\fi

\global\newcount\pnfigno \global\pnfigno=1
\newwrite\ffile
\def\pfig#1#2{Fig.~\the\pnfigno\pnfig#1{#2}}
\def\pnfig#1#2{\xdef#1{Fig. \the\pnfigno}%
\ifnum\pnfigno=1\immediate\openout\ffile=figs.tmp\fi%
\immediate\write\ffile{\noexpand\item{\noexpand#1\ }#2}%
\global\advance\pnfigno by1}

%
%
\def\tfig#1{Fig.~\the\pnfigno\xdef#1{Fig.~\the\pnfigno}\global\advance\pnfigno
by1}

%
%
%
%
\def\figI{y}
\def\ifigure#1#2#3#4{
\midinsert
\ifx\figflag\figI
\vbox to #4truein{
\vfil\centerline{\epsfysize=#4truein\epsfbox{#3}}}
\else
\vbox to .2truein{}
\fi
\narrower\narrower\noindent{\bf #1:} #2
\endinsert
}





\def\M#1#2{\mu_#1^{\phantom{#1}#2}} 



%
%

%


\def\inbar{\,\vrule height1.5ex width.4pt depth0pt}
\def\IB{\relax{\rm I\kern-.18em B}}
\def\IC{\relax\hbox{$\inbar\kern-.3em{\rm C}$}}
\def\ID{\relax{\rm I\kern-.18em D}}
\def\IE{\relax{\rm I\kern-.18em E}}
\def\IF{\relax{\rm I\kern-.18em F}}
\def\IG{\relax\hbox{$\inbar\kern-.3em{\rm G}$}}
\def\IH{\relax{\rm I\kern-.18em H}}
\def\II{\relax{\rm I\kern-.18em I}}
\def\IK{\relax{\rm I\kern-.18em K}}
\def\IL{\relax{\rm I\kern-.18em L}}
\def\IM{\relax{\rm I\kern-.18em M}}
\def\IN{\relax{\rm I\kern-.18em N}}
\def\IO{\relax\hbox{$\inbar\kern-.3em{\rm O}$}}
\def\IP{\relax{\rm I\kern-.18em P}}
\def\IQ{\relax\hbox{$\inbar\kern-.3em{\rm Q}$}}
\def\IR{\relax{\rm I\kern-.18em R}}
\font\cmss=cmss10 \font\cmsss=cmss10 at 10truept
\def\IZ{\relax\ifmmode\mathchoice
{\hbox{\cmss Z\kern-.4em Z}}{\hbox{\cmss Z\kern-.4em Z}}
{\lower.9pt\hbox{\cmsss Z\kern-.36em Z}}
{\lower1.2pt\hbox{\cmsss Z\kern-.36em Z}}\else{\cmss Z\kern-.4em Z}\fi}
\def\IGa{\relax\hbox{${\rm I}\kern-.18em\Gamma$}}
\def\IPi{\relax\hbox{${\rm I}\kern-.18em\Pi$}}
\def\ITh{\relax\hbox{$\inbar\kern-.3em\Theta$}}
\def\IOm{\relax\hbox{$\inbar\kern-3.00pt\Omega$}}



\def\pndate{3/95}

\def\testp{T}

\Title{UPR-651T }{Fluctuating Membranes With Tilt Order}

\centerline{Thomas Powers and Philip Nelson}\smallskip
\centerline{Physics Department, University of Pennsylvania}
\centerline{Philadelphia, PA 19104 USA}
\bigskip\bigskip

Thermal fluctuations are important for amphiphilic bilayer membranes
since typical bending stiffnesses can be a few $k_B T$.  The rod-like
constituent molecules are generically tilted with respect to the local
normal for packing reasons.  We study the effects of fluctuations on
membranes with nematic order, a simplified idealization with the
same novel features as realistic tilt order.
We find that nematic membranes lie in
the same universality class as hexatic membranes, {\it i.e.} the couplings
that distinguish nematic from hexatic order are marginally irrelevant.
Our calculation also illustrates the advantages of conformal gauge,
which  brings great conceptual and technical simplifications compared
to the more popular Monge gauge.

\ifx\prlmode\testp
\fi
\ifx\answ\bigans \else\noblackbox\fi
\Date{\pndate }\noblackbox
\noblackbox


\hfuzz=3truept
\def\pagin#1{}
\lref\DaGuPe{F. David, E. Guitter, and L. Peliti, \cmp{``Critical
properties of fluid membranes with hexatic order,''} J. Phys. France
{\bf48} (1987) 2059.}%
\lref\GuKa{E. Guitter and M. Kardar, \cmp{``Tethering, crumpling, and
melting transitions in hexatic membranes,''}
     Europh. Lett. {\bf13} (1990) 441\pagin{--446}.}
\lref\PaLu{J.-M. Park and T. Lubensky, \cmp{``Kosterlitz-Thouless
transition on fluctuating surfaces,} preprint 1995.}
\lref\bigchiral{P. Nelson and T. Powers, \cmp{``Renormalization of chiral
couplings in tilted bilayer membranes,''} J.  Phys. France II  {\bf3}
(1993) 1535.}
\lref\Conformal{F. David, in \SMMS.}%
\lref\SMMS{See for example {\sl Statistical mechanics of membranes and
surfaces,} D. Nelson {\it et al.}, eds (World Scientific, 1989).}
\lref\PNTRP{P. Nelson and T. Powers, Phys. Rev. Lett. {\bf69} (1992) 3409.}
\lref\MFM{G. Moore and P. Nelson, \cmp{``Measure for moduli,''} Nucl. Phys.
{\bf B266} (1986) 58.}
\lref\Polch{J. Polchinski, \cmp{``Evaluation of the one loop string path
integral,''} Commun. Math. Phys. {\bf 104} (1986) 37.}
\lref\DDK{F. David, ``Conformal field theories coupled to 2-D gravity in the
conformal
gauge,''
 Mod. Phys. Lett. {\bf A3} (1988) 1651; J. Distler and H. Kawai,
``Conformal field theory and 2-D quantum gravity,''  Nucl. Phys. {\bf
B321} (1989) 509.}
\lref\PoStro{J. Polchinski and A. Strominger, ``Effective string theory,''
Phys. Rev. Lett. {\bf67} (1991) 1681.}
\lref\Polbook{A. Polyakov, {\sl Gauge fields and strings}, (Harwood,
1987).}
\lref\DGP{F. David, E. Guitter, and L. Peliti, \cmp{``Critical
properties of fluid membranes with hexatic order,''} J. Phys. France
{\bf48} (1987) 2059.}%
\def\Hexatic{\NePe\DGP\GuKa\PaLu}
\lref\Rigidity{E. Evans and W. Rawicz, Phys. Rev. Lett. {\bf 64} (1990)
2094; M. Bloom, E. Evans, and G. Mouritsen,
Quart. Rev. Biophys. {\bf24} (1991) 293.}
\lref\DNelson{D. Nelson, \cmp{``Defect-mediated phase transitions,''}
in {\sl Phase transitions and critical
phenomena,} vol. 7, ed. C. Domb and J. Lebowitz (Academic Press, 1983).}
\lref\Langhex{R. Kenn, C. B\"ohm, A. Bibo, I. Peterson, H. M\"ohwald,
J. Als-Nielsen, and K. Kjaer, \cmp{``Mesophases and crystalline phases
in fatty acid monolayers,''} J. Phys. Chem., {\bf 95} (1991)
2092\pagin{--2097}.}
\lref\Smhex{M. Cheng, J. Ho, S. Hui, and R. Pindak,
\cmp{``Observation of two-dimensional hexatic behavior in free-standing
liquid crystal films,''} Phys. Rev. Lett. {\bf 61}
(1988) 550.}
\lref\Tilthex{E. Sirota, G. Smith, C. Safinya, R. Plano, N. Clark,
\cmp{``X-ray studies
of aligned, stacked surfactant membranes,''} Science {\bf 242} (1988)
 1406\pagin{--1409}.}
\lref\LanLif{L. Landau and E. Lifshitz, {\sl Theory of elasticity}
(Pergamon Press, 1986).}
\lref\Ripple{
T. Lubensky and F. Mackintosh, \cmp{``Theory of ripple phases of lipid
bilayers,''} Phys. Rev.  Lett. {\bf72} (1993) 1565\pagin{--1568}.}
\lref\Chiral{
P. Nelson and T. Powers, \cmp{``Rigid Chiral Membranes,''} Phys. Rev. Lett.
{\bf69}
(1992) 3409\pagin{--3412}.
}
\lref\Bud{H.-G. D\"obereiner, J. Kas, D. Noppl, I. Sprenger, and E.
Sackmann,
\cmp{``Budding and fission of vesicles,''} Biophys. J. 65 (1993)
1396\pagin{--1403}.}
\lref\rBloom{M. Bloom and O. Mouritsen, \cmp{``The evolution of membranes,''}
in
R. Lipowsky and E. Sackmann, eds., {\sl Biophysics Handbook
on Membranes I: Structure and Conformation} (to be published).}
\let\Bio=\rBloom
\lref\PeLe{L. Peliti and S. Leibler, \cmp{``Effects of thermal fluctuations on
systems with small surface tension,''}
Phys. Rev. Lett. {\bf54} (1985) 1690.} 
\lref\Tether{D.R. Nelson and L. Peliti, \cmp{``Fluctuations in membranes with
crystalline and hexatic order,''} J. Phys. France {\bf 48} (1987)
1085.
}
\lref\NePe{D. Nelson and R. Pelcovits, \cmp{``Momentum-shell recursion
relations, anisotropic spins, and liquid crystal in $2+\epsilon$
dimensions,''} Phys. Rev. B {\bf 16} (1977) 2191.}
\lref\Polyakov{A. Polyakov, \cmp{``Fine structure of strings,''}
Nucl. Phys. {\bf B268} (1986) 406.} 
\lref\WilKog{K. Wilson and J. Kogut, \cmp{``The renormalization group
and the $\epsilon$ expansion,''} Phys. Reports {\bf C12}
(1974) 75\pagin{--200}.}
\def\Measure{\DDK} 
\lref\DDK{F. David, ``Conformal field theories coupled to 2-D gravity in the
conformal
gauge,''
 Mod. Phys. Lett. {\bf A3} (1988) 1651; J. Distler and H. Kawai,
``Conformal field theory and 2-D quantum gravity,''  Nucl. Phys. {\bf
B321} (1989) 509.}
\lref\Gauge{??SHIFT SYMMETRY}
\lref\PePr{L. Peliti and J. Prost, \cmp{``Fluctuations in membranes with
reduced
symmetry,''} J. Phys. France {\bf 50} (1989) 1557.}  


The study of fluctuating surfaces in an intriguing statistical
mechanics problem with numerous applications.  Amphiphilic molecules
in water can self-assemble into
thin flexible bilayer membranes which provide an experimental realization
of random surfaces \SMMS.  As we recall below, a membrane's internal
order determines the nature of its shape fluctuations.  Although fluctuating
membranes with hexatic bond-orientational order have received a lot of
theoretical attention \Hexatic, the generic situation is for the rod-like
constituent molecules to {\it tilt} with respect to the local surface normal.
We
shall see that tilt order differs from hexatic order since tilt allows certain
anisotropic couplings.  In this letter we consider ``nematic''
membranes, the simplest
membrane model with in-plane orientational order and anisotropy.  We find
the anisotropy to be marginally irrelevant; these membranes may be considered
hexatic at very long length scales.

Our framework is continuum elastic theory since the micron size of
membrane structures is much larger than the typical molecular size, a
few tens of {\AA}ngstroms.  If the molecules are allowed to take up their
preferred area, then surface tension will be unimportant and bending
rigidity will dominate.  Typical bending moduli for lipid bilayer
membranes are on the order of $10 k_B T$ \Rigidity, which is low
enough for thermal fluctuations to play a role.  We study
fluid-like membranes in which the molecules are free to diffuse
along the membrane in response to shape fluctuations.  Therefore
shape and orientational order are the only elastic degrees of freedom,
and our elastic theory must be coordinate-invariant \Conformal.

The simplest type of in-plane orientational order is hexatic order
\DNelson.  Pure hexatic order is readily seen in Langmuir monolayers
\Langhex\ and thin smectic liquid crystal films \Smhex, but this order has
not to our knowledge been confirmed in bilayer membranes.  Instead,
in the ordered phases the
constituent molecules generically tilt for packing reasons.  If tilt
and hexatic order are present together \Tilthex, then in the absence
of defects we can for the purposes
of elastic theory treat these as locked together.  Tilt and hexatic order
break the same rotational symmetry, and together lead to just one elastic
mode.  Since tilt order has less symmetry than hexatic order (see below),
the relevant order is tilt when both tilt and hexatic order are present.

One can represent hexatic order on a membrane by
a unit tangent vector field \DGP.  Each vector points from a molecule
to one of the molecule's nearest neighbors and is defined up to $2 \pi/6$
rotations.  Demanding this symmetry automatically brings along a
larger symmetry,
{\it i.e.} global rotations through an arbitrary angle; in this sense
hexatic order is isotropic.  This is reminiscent
of the isotropy of the continuum elastic theory of a two-dimensional
triangular lattice \LanLif, but holds even on curved surfaces, whose
principal curvatures could in principle have cared about a global rotation
of in-plane order.

Similarly one can represent tilt order as a unit tangent
vector field $\hat m$ defined by the direction of the projection of the
axis of each molecule on the local tangent plane.  We consider only
the elastic modes so we fix the polar angle between the normal $\hat n$
and the
molecular axis.  Bilayers lack a preferred normal, so to complete
our specification of $\hat m$ we demand the bilayer symmetry $\hat n
\rightarrow
-\hat n$, $\hat m \rightarrow -\hat m$ \PePr\bigchiral.
In contrast to the hexatic case,
the energy for bending a membrane along an axis
parallel to the direction of tilt can clearly be different from the energy to
bend
the membrane the other way.
(Mathematically we will see in a moment how this conclusion arises from the
reduced symmetry of tilt order.)
This tilt-shape coupling has a number of simple consequences; for example,
tilt order is responsible for the $P_{\beta'}$ ripple
phases \Tilthex\Ripple.  Tilt order is also manifested dramatically when the
constituent
molecules are chiral:  the microscopic molecular chirality affects
macroscopic membrane shape only in the presence of tilt \Chiral.
Tilt order can play a role in the budding of artificially manufactured
vesicles \Bud\ and it may be the typical order in biological membranes \Bio.

The nature of membrane shape fluctuations depends on the degree of internal
order.  We characterize these fluctuations by using the Wilson renormalization
group to compute the long-distance behavior of the elastic couplings.  For
example, the bending rigidity of a fluid membrane becomes ineffective
beyond a persistence length $\xi_P$, leading to a crumpled phase at
any nonzero temperature \PeLe.\foot{We ignore self-avoidance since we will
ultimately study a fixed point at large stiffness.}  Internal
order tends to stiffen a membrane.  For stiff enough elastic constants,
the bending rigidity for a hexatic membrane reaches a fixed point at
very long length scales, possibly leading to a ``crinkled'' phase with
quasi-long-range order in the normals \DGP.  Finally, non-self avoiding
tethered membranes (crystalline membranes with an infinite core energy
for dislocations) undergo a crumpling transition and have a flat phase
at low temperature \Tether.  It is natural to extend these analyses to
the case of tilt order, since as we remarked it is experimentally the most
relevant regime with orientational order.
To keep our formulas compact we will impose a ``nematic,'' or 2-atic
symmetry to get the simplest model with anisotropy; the same technique
applies to the more realistic case with no discrete symmetry.
We will then ask,
does the anisotropy lead to new fixed points and new physics, or is the
anisotropy marginally irrelevant as in flat thin liquid crystal films \NePe?

To answer these questions with precise calculations, we must choose
coordinates.  The most popular choice for such calculations is Monge
gauge, in which the surface is parametrized by its height above some
flat reference surface \PeLe\bigchiral. 
It turns out that Monge gauge is not very convenient for tilted membranes;
conformal gauge, as used for example by Polyakov \Polyakov,
is much better-suited.
Conformal gauge has the technical advantages of leading to compact
expressions and thus less algebra than Monge gauge, as well as
the conceptual advantage of making explicit
a useful separation between extrinsic and intrinsic geometry.

We begin by describing the geometrical constructions necessary to write
down the tilt free energy.  Even with the benefits of conformal gauge,
the formulas can get a bit long. 
Thus, we make some inessential but simplifying assumptions.  Next, we
briefly summarize the issues involved in conformal gauge calculations.
Finally, we present our result and discuss its implications.

Following the notation of \Polyakov, we parametrize our surface
by $\vec x(\xi)$, where $\vec x$ is a three dimensional vector and
$(\xi^1,\xi^2)$ are the (arbitrary) two dimensional coordinates.  From
$\vec x(\xi)$ we construct the metric tensor $g_{ab}=\pa_a\vec x\cdot
\pa_b\vec x$ and the second fundamental form $K_{ab}=\hat n \cdot\pa_a
\pa_b\vec x$, where $\pa_a = \pa/\pa\xi^a$
and $\hat n$ is the local normal. We denote the covariant derivative
associated with $g_{ab}$ by $\nabla$, the inverse of $g_{ab}$ by
$g^{ab}$, and the determinant of $g_{ab}$ by $g$.  It is useful
to introduce a local set of orthonormal frames $\hat e_\alpha$; we denote
by $e_\alpha{}^a$ the change-of-basis matrix that converts orthonormal
frame indices $\alpha, \beta, ...$ to coordinate indices $a,b, ....$
Thus for example $\nabla_a m^\alpha = \pa_a m^\alpha + \epsilon^\alpha{}_\beta
\Omega_a m^\beta$ where $\Omega_a$ is the spin connection \DGP.

The elastic free energy for tilted bilayer membranes must have coordinate
invariance, Euclidean invariance, and the discrete bilayer symmetry discussed
above.  Writing only the bulk terms quadratic in the curvature or derivatives
of $\hat m$, {\it i.e.} only the marginal terms, we find $F_{\rm tilt}=F_1+F_2$
where
\eqnn\nematic
$$
\eqalignno{F_1 =&
{1\over 2}\int\dd^2\xi\sqrt{g}[
k_1(\nabla\cdot \hat m)^2
+k_2(\nabla\times\hat m)^2+\kappa(K_a{}^a)^2\cr
&\qquad+\kappa_1\hat m\cdot K\cdot K\cdot\hat m
+\kappa_2(\hat m\cdot K\cdot\hat m)^2 ]\cr
F_2 = & {1\over 2}\int\dd^2\xi\sqrt{g}
[\beta_1 (\hat m\cdot K\cdot\hat m)(\nabla\cdot\hat m) +
\beta_2(K_a{}^a)(\nabla\cdot \hat m)\cr
&\qquad+\beta_3 K^a{}_b\nabla_am^b+
\beta_4 m^c K_{c a}m^b\nabla_b m^a]&\nematic. }
$$
In \nematic, we contract indices with $g^{ab}$, and $\nabla\times\hat m =
\epsilon^a{}_b\nabla_a m^b$, where $\epsilon_{ab}$ is the covariant
antisymmetric tensor.\foot{In \nematic\ we have corrected a redundancy in
\bigchiral; we note here that the $\alpha_1$ and $\alpha_2$ terms
of \bigchiral\
differ by a total derivative.}  Note that $F_1$ reduces to the hexatic membrane
free energy when $\kappa_1=\kappa_2=0$ and $k_1=k_2=k_A$, since even
on a curved surface $(\nabla\cdot\hat m)^2+(\nabla\times\hat m)^2=
\nabla_a m^b \nabla^a m_b.$  Following Nelson and Pelcovits \NePe,
we find it convenient to take $k_2>k_1$ without loss of generality
and rewrite the terms involving only $\hat m$ and the metric as
$k_2 \nabla_a m^b \nabla^a m_b + \bar k (\nabla \cdot \hat m)^2$, where
$\bar k = k_2 - k_1$.

As we alluded to above, for large stiffnesses $\kappa, k_A$, hexatic
membranes are governed by a line of fixed points in the
$\kappa\inv$--$k_A{}\inv$ plane \DGP.  Our first simplification for tilt is to
study
the stability of this line against anisotropy by working to first
order in the anisotropic couplings.  We further simplify our job by
dropping the terms of $F_2$ in \nematic.  This truncation is mathematically
consistent because $F_1$ is a complete list of terms with the ``nematic''
$\hat m\rightarrow -\hat m$ symmetry.    We do not know of an
experimental system with this symmetry.  One can imagine a membrane made
of rod-like molecules that lie parallel to the local tangent plane, or
perhaps more realistically a membrane with stiff rod-like molecules
aligned along the local normal but with a rectangular cross-section.
In any case, this free energy is the simplest membrane
model with in-plane order and anisotropy; we expect it to be
qualitatively similar to the case of real tilt.

We study the long-distance behavior of our model with the Wilson
momentum-shell renormalization group \WilKog.  The first step of this
procedure is to decompose the fields into slowly varying and rapidly
varying parts $\vec x = \vec x_0 + \vec x_1, \hat m = \vec m_0 + \vec m_1$,
where $\vec x_0, \vec m_0$ have Fourier modes with wavevector $k$
satisfying $|k|<\Lambda/b$, and $\vec x_1, \vec m_1$ have
Fourier modes in a shell $\Lambda/b<|k|<\Lambda$.  $\Lambda=2\pi/a$ is the
wavevector cutoff corresponding to the short-distance cutoff $a$, and
$b$ is a number slightly greater than one.  The next step is to
coarse-grain the system by tracing over the fast modes in the partition
function.  We write the partition function for the membrane as a path
integral over the shape and orientational degrees of freedom \SMMS.
Since membranes are stiff but not completely rigid, we work to first order
in $T/\kappa$.\foot{Since we work in the stiff regime, we ignore defects
in the in-plane order.  These will not qualitatively change our result
(but see \PaLu).}
This approximation amounts to integrating out the
fast modes in the Gaussian approximation.  Finally, we rescale the
distances to restore the cutoff to its original value and obtain the
recursion relations for the effective couplings.

To carry out these calculations, we must choose coordinates.  The metric
$g_{ab}=\pa_a\vec x\cdot\pa_b\vec x$ at each point of a two dimensional
surface is a symmetric $2\times2$ matrix and thus has three independent
degrees of freedom.  Two of these are removed by the two independent
coordinate degrees of freedom, leaving one physical degree of freedom.
In fact, we can always find local coordinates in which the metric
is a spatially dependent conformal factor times the trivial metric \Conformal:
\eqn\conformal{g_{ab}(\xi)=\rho(\xi)\delta_{ab}.}
Conformal gauge is the two dimensional analog of arc-length parameterization
of a curve in space.

The tensor fields associated with a surface have a compact form in conformal
gauge.  For example, the mean curvature is $\hat n\cdot \nabla^2\vec x
= \rho^{-1}\hat n\cdot \pa^2\vec x$, the change of basis matrix is
$e_a{}^\alpha=\sqrt\rho \delta_{a\alpha}$, and the spin connection is
$\Omega_a = {1\over 2}\epsilon_{ab}\partial_b\log \rho$ \Conformal.

For purposes of illustration we first consider the {\it fluid} membrane
(no in-plane order) in
conformal gauge, as discussed by Polyakov \Polyakov.  Since $\rho$ depends
on $\vec x$, the fluid membrane free energy is a complicated
nonlinear functional of $\vec x$ and not in the most convenient form
to carry out the renormalization group procedure.  To treat $\rho$ as
an independent field, we introduce a delta function constraint and
then enforce the constraint by introducing some Lagrange
multiplier fields \Polyakov:
\eqnn\lagrange
$$
\eqalignno{
Z=&
\int[{\rm d}\vec x] \exp\left[-{\kappa\over {2T}}\int{\rm d}^2\xi
\rho^{-1}(\pa^2\vec x)^2\right]\cr
=&\int[{\rm d}\vec x][{\rm d}\rho]\delta[\pa_a\vec x\cdot\pa_b\vec x - \rho
\delta_{ab}]\exp\left[-{\kappa\over {2T}}\int{\rm d}^2\xi
\rho^{-1}(\pa^2\vec x)^2\right]\cr
=&\int[{\rm d}\vec x][{\rm d}\rho][{\rm d}\lambda^{ab}]e^{ -F} & \lagrange}
$$
where
\eqn\fluid{F={\kappa\over {2T}}\int{\rm d}^2\xi[\rho^{-1}(\pa^2\vec x)^2
+i \lambda^{ab}(\pa_a \vec x\cdot \pa_b \vec x- \rho \delta_{ab})].}
To complete the specification of the partition function, we must define
the functional measure.  There are various factors that
arise from coordinate invariance
\Measure.  We simply note that the geometrical measure factors will not
enter our calculations to $O(T/\kappa)$, and the Liouville counterterm
will not enter the recursion relations for the bending stiffnesses.  A
further advantage of conformal gauge over Monge gauge is that these
conformal gauge measure factors have been well-studied \Measure, making
conformal gauge the better choice for calculations beyond $O(T/\kappa)$.

The Wilson renormalization group procedure requires the most general low
order expression consistent with all the symmetries; \fluid\ is not obviously
of this form.  The Lagrange multiplier field enters in a very specific way.
More precisely, since the renormalization group is an iterative procedure,
we must be sure that after eliminating short wavelength modes the
long-wavelength effective free energy differs from the original only
by the values of the couplings.  If {\it e.g.} a $\lambda^2$ term
were generated, then the delta function constraint would be softened
to a Gaussian and we could not define the recursion relations.  
Such terms do in fact appear, but always suppressed by powers of the
short-distance cutoff.

The extra fields $\lambda^{ab}$ and $\rho$ were supposedly introduced to
simplify the calculation; it is natural to ask why the extra fields
do not lead to more complexity.  The continued sharpness of the delta
function during coarse-graining leads to a technical simplification.
To see this, we recall Polyakov's trick to diagonalize the fluctuation
part of the free energy functional \Polyakov.  He decomposes the rapidly
varying part $\lambda^{ab}_1$ of the Lagrange multiplier field
$\lambda^{ab}$ into a traceless and transverse trace part taking
\eqn\decomp{\lambda^{ab}_1=\pa_a f_{1b}+\pa_b f_{1a}-\delta_{ab}
\pa_c f_{1c} + \left(\delta_{ab}-{\pa_a\pa_b\over{\pa^2}}\right)\zeta_1.}
Using this decomposition, we expand the free energy to quadratic
order in the fast fields and split its quadratic part $F^{II}$
into a free part and a part
to be treated as a perturbation $F^{II}=F_A+F_B$,
\eqn\FA{F_A={\kappa\over{2T}}\int{\rm d}^2\xi[\rho_0{}^{-1}(\pa^2\vec x_1)^2
-2 i (\pa^2f_{1a})(\pa_a\vec x_0)\cdot\vec x_1 - i \zeta_1 \rho_1],}
\eqnn\FB
$$
\eqalignno{F_B=&{\kappa\over{2T}}
\int{\rm d}^2\xi[i\lambda^{ab}_0(\pa_a \vec x_1)
\cdot(\pa_b \vec x_1) - 2\rho_1\rho_0{}^{-2}(\pa^2\vec x_0)\cdot(\pa^2\vec
x_1)\cr
&+\rho_1{}^2\rho_0{}^{-3}(\pa^2\vec x_0)^2-2 i
\lambda^{ab}_1(\pa_a\pa_b\vec
x_0)\cdot \vec x_1].}
$$
In the diagrammatic expansion of the effective free energy, $F_A$
determines the propagators and $F_B$ determines the vertices.  Since
there is no $\zeta_1{}^2$ term in \FA, the $\rho_1$ propagator
is zero: $\langle \rho_1 \rho_1\rangle = 0$.  Many diagrams that otherwise
would have contributed are thus zero.  Also, the bending free energy for
a fluid membrane is independent of derivatives of $\rho$, so we can
treat $\rho_0$  as a constant.  Thus the conformal factor $\rho$ does
not play much of a role in the calculations.

Turning to our model, we introduce another Lagrange multiplier $\mu$
to enforce $\hat m^2=1$.  Introducing $\mu$ is more convenient than
parametrizing $\hat m$ as $m^\alpha=(\cos \theta, \sin\theta)^\alpha$,
since all the terms of \nematic\ are nonlinear in $\theta$.  Working
directly in terms of $\hat m$ is covariant and makes for compact
expressions.  Also, the Lagrange multiplier trick takes care of
field renormalization automatically ({\it cf.} \NePe).

We are now ready to outline our calculation.  Denoting the fields
$(\vec x, \rho, f,\zeta, \vec m, \mu)$ collectively as $\phi$, we
want to compute the effective free energy
\eqn\effenergy{F_{\rm eff}[\phi_0]=-\log\int[{\rm
d}\phi_1]\exp\left(-\int\phi_1
{\cal O}[\phi_0]\phi_1\right)={1\over2}{\rm Tr}\log{\cal O}[\phi_0],}
where ${\cal O}[\phi_0]$ is the matrix associated with the fluctuation
free energy $F^{II}={1\over2}\int{\rm d}^2\xi\phi_1{\cal O}[\phi_0]\phi_1$.
We split ${\cal O}[\phi_0]$ into a free part ${\cal O}_A[\phi_0]$ and
an interaction part ${\cal O}_B[\phi_0]$.  Since we work to first
order in the anisotropic couplings $\bar k, \kappa_1, \kappa_2$, we
treat all the subterms in the expansion of the anisotropic terms
as perturbations.  Thus,
\eqn\free{
{\cal O}_A=\pmatrix{\rho_0\inv\pa^4\delta_{ab}&-i(\pa_b\vec x_0)\pa^2& & & &
\cr
-i\pa^2(\pa_a\vec x_0)&0& & & & \cr  & &0&-{i\over2}& & \cr
 & &-{i\over2}&-{k_2\over 4}\rho_0^{-2}\pa^2& & \cr & & &
&-k_2\pa^2\delta_{\alpha\beta}&k_2 m_0^\beta\cr
& & & &k_2 m_0^\alpha&0}.}

Here we can see that the block diagonal form of ${\cal O}_A$ (and
of ${\cal O}_A{}^{-1}$, the matrix of propagators) leads to a clear
separation between extrinsic and intrinsic geometry.  There are no
propagators connecting $\vec x_1$ fields (extrinsic) with $\rho_1$
or $\vec m_1$ fields (intrinsic).  In our expansion, the only coupling
between extrinsic and intrinsic geometry is through graph vertices
derived from the Lagrange multiplier term of \fluid.

The free energy is given by expanding the logarithm in \effenergy\
in powers of ${\cal O}_A{}^{-1}{\cal O}_B$:
\eqn\expand{{1\over2}{\rm Tr}\log({\cal O}_A+{\cal O}_B)={1\over2}
{\rm Tr}[\log{\cal O}_A+{\cal O}_A\inv{\cal O}_B-
{1\over2}{\cal O}_A\inv{\cal O}_B{\cal O}_A\inv{\cal O}_B+\cdot\cdot\cdot]
.}
The first term of \expand\ corresponds to graphs with no vertices, the
second to graphs with one vertex, and so on.  Since ${\cal O}_A$ depends
on the slow fields we must keep the first term; unlike in Monge gauge,
graphs with no vertices contribute to the renormalization of the stiffnesses.
Evaluating the traces of \expand\ leads to the effective free energy
\eqnn\eff
$$
\eqalignno{& F_{\rm eff}=
{1\over{2T}}\left[\kappa+{T\over{2\pi\kappa}}\left(-\kappa+
{3 \over 8} k_2+
{5\over 16}\bar k+{17 \over 4}\kappa_1-{9\over 32}\kappa_2\right)\log b\right]
\int\dd^2\xi \rho_0^{-1}(\pa^2\vec x_0)^2\cr
&+{1\over{2T}}\left[\kappa_1+ {T\over{8\pi\kappa}}\left(-{\bar k}-7\kappa_1
+11\kappa_2\right)\log b\right]\int\dd^2\xi
\rho_0^{-1}m_0^\alpha(\pa_\alpha\pa_a\vec x_0)\cdot
(\pa_a\pa_\beta\vec x_0)m_0^\beta\cr
&+{1\over{2T}}\left[\kappa_2 -{{T\kappa_2}\over{8\pi\kappa}}\log
b\right]\int\dd^2\xi \rho_0^{-1}
m_0^\alpha m_0^\beta m_0^\gamma m_0^\delta(\pa_\alpha\pa_\beta\vec
x_0)\cdot(\pa_\gamma\pa_\delta
\vec x_0)\cr
+{1\over{2T}}\int&\dd^2\xi{ } i \lambda_0^{ab}\left[\pa_a\vec x_0\cdot\pa_b
\vec x_0
-\rho_0\left(\delta_{ab}+{T\over{4\pi\kappa}}\bigl(-\delta_{ab}
+\delta_{ab}{{2\kappa_1+\kappa_2}\over{8\kappa}}
-m_{0a} m_{0b}{{\kappa_1+\kappa_2}\over{2\kappa}}\bigr)\right)\log b\right]\cr
&+{k_2\over{2T}}\int\dd^2\xi(\pa_\alpha m_0^\beta)^2\left(1+{\bar k T\over{2\pi
k_2}}\log b\right)
+{\bar k\over {2T}}\int\dd^2\xi(\pa_\alpha m_0^\alpha)^2\left(1-{T\over{\pi
k_2}}\log b\right)
.&\eff}
$$
In deriving \eff, we have used $\pa_a\vec x_0\cdot \pa_b\vec x_0 =
\rho_0 \delta_{ab}+O(T/\kappa)$ and $\vec m_0^2 = 1 + O(T/\kappa)$.

The purpose of quoting the long expression \eff\ is to point out
an unexpected term:  the term with $\lambda^{ab} m_{0a} m_{0b}$
did not appear in the original free energy.  This term spoils
the conformal gauge condition $\pa_a\vec x_0\cdot\pa_b\vec x_0 = \rho_0
\delta_{ab}$ for the slow fields, but it does not spoil the renormalization
group calculation.  We simply integrate over $\lambda_0^{ab}$ to
get a modified delta function relating $\pa_a\vec x_0\cdot\pa_b\vec x_0$,
$\rho_0\delta_{ab}$, and $m_{0a}m_{0b}$.  Using the delta function to
integrate over $\rho_0$, we obtain a free energy of the same form
as \nematic\ but with modified coefficients.  Substituting the hexatic
fixed line relation $k_2 = 4\kappa$, we find the linearized recursion
relations ${\rm d}\vec g/{\rm d}\log b = M \vec g$, with
\eqn\M{M={1\over{4\pi\beta\kappa}}\pmatrix{-{3\over{\beta\kappa}}&{12\over{\beta\kappa}}&
{5\over8}&{17\over2}&{-13\over16}\cr
 & &-2&1&1\cr
 & &-1&-2&-2\cr
 & &-{1\over2}&-{7\over2}&{13\over2}\cr
 & & & &-{3\over2}}
,}
$\vec g = ((\beta\kappa)^{-1},(\beta k_2)^{-1},\beta\bar k, \beta\kappa_1,
\beta\kappa_2)$, and $\beta = 1/T$.  The matrix $M$ has
eigenvalues $-3/(4\pi\beta^2\kappa^2)$ and $0$ corresponding to the
hexatic membrane,
and three more negative eigenvalues, $(-9\pm\sqrt41)/(16\pi\beta\kappa),
-3/(8\pi\beta\kappa)$, corresponding to the anisotropic couplings.

The hexatic fixed line is therefore stable to the anisotropic terms obeying
the $\hat m \rightarrow - \hat m$ symmetry.  Beyond a length scale exponential
in the stiffness $\kappa/T$ we expect nematic membranes to behave as
hexatic membranes.  This is in accord with our intuition that since there
is no true long-range in-plane order in this two dimensional system,
at very long length scales anisotropy should be irrelevant.  The exponential
dependence of these length scales on the stiffness means in practice
anisotropic membranes will differ from their hexatic counterparts.  We
remark that Peliti and Prost have also argued that anisotropic membranes
should lie in the same universality class as hexatic membranes \PePr.
However our calculation shows that the actual way in which this happens
is different from the screening mechanism they proposed.

We have seen that conformal gauge is well-suited for determining the
long-distance properties of fluctuating membranes with in-plane order.
We can easily include the terms we ignored to extend our results
to more realistic tilted membranes.
Our technique is useful for other applications.  One example is the critical
exponents  of a
crinkled tilted membrane; we expect these exponents to differ from
the hexatic case since the anisotropy is only marginally irrelevant ({\it cf.}
\DGP).


{\frenchspacing
\ifx\prlmode\testp\else\vskip1truein \leftline{\bf Acknowledgements}
\noindent\fi
We would like to thank
Y. Hatwalne,
R. Kamien,
T. Lubensky,
J. Park,
L. Peliti,
A. Polyakov,
and J. Selinger for useful discussions.  We are especially indebted to
T. Lubensky for pointing out an error in an early version of the calculation.
PN acknowledges NSF
grants PHY88-57200
and the Donors of the Petroleum Research Fund,
administered by the American Chemical Society, for the partial support
of this research.}

\listrefs
\bye